\documentclass{article}

\usepackage[final]{neurips_2021_ml4ps}
\usepackage{aas_macros} 
\usepackage{subcaption}

\usepackage[utf8]{inputenc} 
\usepackage[T1]{fontenc}    
\usepackage{hyperref}       
\usepackage{url}            
\usepackage{booktabs}       
\usepackage{amsfonts}       
\usepackage{nicefrac}       
\usepackage{microtype}      
\usepackage{xcolor}         
\usepackage{mathtools} 
\usepackage{dsfont} 
\title{Re-calibrating Photometric Redshift Probability Distributions Using Feature-space Regression}

\author{%
  Biprateep Dey, Jeffrey A. Newman, Brett H. Andrews  \\
  Dept. of Physics and Astronomy and PITT-PACC\\
  University of Pittsburgh, Pittsburgh, PA 15260, USA \\
  \texttt{\{biprateep, janewman, andrewsb\}@pitt.edu} \\

\And
  Rafael Izbicki \\
  Dept. of Statistics\\
  Federal University of S\~ao Carlos (UFSCar)\\
  São Carlos, Brazil \\
  
  \And
  Ann B. Lee, David Zhao\\
  Dept. of Statistics \& Data Science \\
  Carnegie Mellon University \\
  Pittsburgh, PA 15213, USA \\

  \And
  Markus Michael Rau \\
  Dept. of Physics \& McWilliams Center for Cosmology \\
  Carnegie Mellon University, Pittsburgh, PA 15213, USA\\  
  \&  \\
  High Energy Physics Division\\
  Argonne National Laboratory, Lemont, IL 60439, USA \\
  \And 
  Alex I. Malz \\
  Ruhr-University Bochum, Astronomical Institute \\
  German Centre for Cosmological Lensing\\
  Universitätsstr. 150, 44801 Bochum, Germany \\
}

\begin{document}

\maketitle

\begin{abstract}
Many astrophysical analyses depend on estimates of redshifts (a proxy for distance) determined from photometric (i.e., imaging) data alone. Inaccurate estimates of photometric redshift uncertainties can result in large systematic errors. However, probability distribution outputs from many  photometric redshift methods do not follow the frequentist definition of a Probability Density Function (PDF) for redshift --- i.e., the fraction of times the true redshift falls between two limits $z_{1}$ and $z_{2}$ should be equal to the integral of the PDF between these limits. Previous works have used the global distribution of Probability Integral Transform (PIT) values to re-calibrate PDFs, but offsetting inaccuracies in different regions of feature space can conspire to limit the efficacy of the method. We leverage a recently developed regression technique that characterizes the local PIT distribution at any location in feature space to perform a local re-calibration of photometric redshift PDFs. Though we focus on an example from astrophysics, our method can produce PDFs which are calibrated at all locations in feature space  for any use case.\footnote{Code available at: \url{https://github.com/biprateep/recalibrate-pdfs}}
\end{abstract}

\section{Introduction}
Galaxy distance, as measured by redshift, is essential for estimating intrinsic luminosity and 3D location in space, which is crucial information for many astrophysical studies.  High-precision redshifts require resource-intensive observations and will only be feasible for a few percent of galaxies in upcoming photometric surveys.  Thus, photometric redshifts (photo-$z$'s)---redshifts estimated from imaging alone---will be necessary.  Furthermore, accurate photo-$z$'s are critical for some science cases (e.g., weak lensing cosmology), but PDFs from both main methods of photo-$z$ estimation (galaxy spectral template-based and machine learning-based) fail to satisfy the frequentist definition of a PDF for redshift \citep{Dahlen2013CANDELSPhotoz,Kodra2019,Schmidt2020PhotozComparison}.  The fraction of times the true redshift falls between two limits $z_{1}$ and $z_{2}$ should equal the integral of a properly-defined PDF between these limits, for any arbitrary subset of the test data.

 Current metrics used to measure the quality of calibration, like the distribution of the values of the cumulative distribution function (CDF) evaluated at the true redshift of the object (the Probability Integral Transform or PIT; see Eq.~\ref{eq:PITdef}) can favor pathological but un-informative PDFs \citep{Schmidt2020PhotozComparison}. Moreover, overall uniformity of PIT values is possible even if particular subsets of the same test data are poorly-calibrated \citep{Zhao2021CDEDiagnostics}. If the PDFs are well-calibrated, then the distribution of the PIT values of a test sample will be uniform between 0 and 1 or their corresponding CDF will follow the identity line for any arbitrary subset of the test data. The same can be visualized with a P-P plot that shows the empirically calculated CDF versus their theoretical expected values. Ideally, the P-P plot should closely follow the identity line, but it often does not.
 
 Several previous works have studied PDF re-calibration (e.g., \citealt{Niculescu2005Re-calibration, Rau2015Re-calibration, Kuleshov2018Re-calibration}), though none can ensure that PDFs are well-calibrated at every point in feature space. \citet{Bordoloi2010PhotozCalibration} described a method to re-calibrate PDFs using a single correction factor based on the overall distribution of PIT values, which ensures a uniform global distribution of PIT values, but this single correction factor is applied to all PDFs and does not account for local variations. Importantly, these local inconsistencies in feature space can be detected using tests like the ones proposed in \citet{JitkrittumK2020LocalTest} and \citet{Zhao2021CDEDiagnostics}, which we will leverage in our method.

In this work, we develop a local PDF re-calibration procedure that uses an estimate of the local distribution of PIT values (from \citet{Zhao2021CDEDiagnostics}) to calculate a correction factor at any location in feature space. As a proof-of-concept, we train a model to predict photo-$z$ PDFs using galaxy magnitudes and colors, which are measures of the amount of light detected in broad wavelength regions. We use the FlexZBoost \citep{Izbicki2017FlexzBoost,Dalmasso2020FlexZBoost} algorithm, which was demonstrated as the best performing photo-$z$ prediction algorithm among the ones compared by \citet{Schmidt2020PhotozComparison}, though any machine learning algorithm producing PDFs will suffice. FlexZBoost uses gradient boosted decision trees (specifically XGBoost; \citealt{Chen2016XGBoost}) to predict photo-$z$ PDFs by minimizing the Conditional Density Estimate (CDE) loss. We use the TEDDY data sets \citep{Beck2017TeddyHappy} to train and test our methods. The TEDDY data set is divided into four subsets which provide us with a test bed for photo-$z$ algorithms to train and test on various distributions of the feature space. We train FlexZBoost on a random 70\% subset of the TEDDY-A data set and use the remaining 30\% as a calibration/validation set. We take the initial photo-$z$ PDFs produced by FlexZBoost as our starting point, which we re-calibrate using 30\% of TEDDY-A as our calibration set and test our methods on TEDDY-B and C data sets. TEDDY-B has the same distribution of features as TEDDY-A, whereas TEDDY-C has a slightly different distribution but spans the same input space.

\section{Re-calibration Procedure}
Let $\widehat{p}(z|\mathbf{x})$ be the initial estimate of the true PDF $p(z|\mathbf{x})$ of the target variable $z$ (redshift) given the input features $\mathbf{x}$ (galaxy colors and magnitudes). The random variable corresponding to $z$ is denoted by $Z$. We define the local Probability Integral Transform (PIT) corresponding to this initial estimate as:
\begin{equation}
    \widehat{\mathrm{PIT}}(z,\mathbf{x}) = \int_{0}^{z} \widehat{p}(z'|\mathbf{x})dz' = \widehat{F}(z|\mathbf{x}) \label{eq:PITdef}
\end{equation}
where $\widehat{F}$ is the cumulative distribution function associated with $\widehat{p}$. Using a labeled calibration set (30\% of TEDDY-A) and a suitable regression method (XGBoost; \citealt{Chen2016XGBoost} in our case), we estimate the CDF of PIT values as a function of $\mathbf{x}$ following the method described in \citet{Zhao2021CDEDiagnostics}. We perform a separate regression for each value of $\alpha$ on a pre-chosen grid $\mathbb{G} \subset [0,1]$ of coverage levels to get the CDF of PIT values ($r^{\widehat{p}}_{\alpha}(\mathbf{x})$):
\begin{equation}
    r^{\widehat{p}}_{\alpha}(\mathbf{x}) \coloneqq \mathds{P} \left( \widehat{\mathrm{PIT}}(Z,\mathbf{x})\leq \alpha | \mathbf{x} \right) = \mathds{P}\left( Z\leq \widehat{F}^{-1}(\alpha|\mathbf{x})|\mathbf{x} \right) \label{eq:ralphadef}
\end{equation}

 If our initial PDFs are locally calibrated, then the relation $r^{\widehat{p}}_{\alpha} = \alpha $ should hold for any $\mathbf{x}$, i.e., a plot of $r^{\widehat{p}}_{\alpha}$ vs. $\alpha$ (also called Amortized Local P-P plots or ALP plots) should closely follow the identity line. Most photo-$z$ estimators do not produce locally calibrated PDFs and this relation often does not often hold (see e.g., Fig.~\ref{fig:local}). 

To re-calibrate the original PDF estimates $\widehat{p}(z|\mathbf{x})$ such that the relation $r^{\Tilde{p}}_{\alpha} \approx \alpha $ holds for the new PDFs, $\Tilde{p}$, for a new unseen test data set, we define, $\beta \coloneqq r^{\widehat{p}}_{\alpha}$ for each $\alpha \in \mathbb{G}$ and define a new cumulative distribution function, $\Tilde{F}$, such that:
\begin{equation}
    \Tilde{F}^{-1} \left( \beta|\mathbf{x} \right) = \widehat{F}^{-1} \left( \alpha|\mathbf{x} \right) \label{eq:inverseCDF}
\end{equation}

Then by construction the new PDFs, $\Tilde{p}$, will be calibrated since
\begin{equation}
    r^{\Tilde{p}}_{\beta}(\mathbf{x}) = \mathds{P}\left( Z \leq \Tilde{F}^{-1}(\beta | \mathbf{x})|\mathbf{x}\right) = r^{\widehat{p}}_{\alpha} = \beta
\end{equation}
Now, for $\Tilde{z}=\Tilde{F}^{-1} \left( \beta|\mathbf{x} \right)$ we will have,

\begin{align}
        \int_{0}^{\Tilde{z}} \Tilde{p}(z'|\mathbf{x})dz' &= \beta \\
    \implies \Tilde{p}(\Tilde{z}|\mathbf{x}) - \Tilde{p}(0|\mathbf{x}) &= \frac{d\beta}{dz'} = \frac{d\beta}{d\alpha}.\frac{d\alpha}{dz'} 
\end{align}
Eqs.~\ref{eq:inverseCDF} and \ref{eq:PITdef} imply that $\Tilde{p}(\Tilde{z}|\mathbf{x})=\Tilde{p}(z|\mathbf{x})$ and Eq.~\ref{eq:ralphadef} implies that $\frac{d\alpha}{dz'}=\widehat{p}(z|\mathbf{x})$. It is not physical to have any object at redshift $0$ so we can assume $\Tilde{p}(0|\mathbf{x})=0$. This gives us the relation:
\begin{equation}
    \Tilde{p}(z|\mathbf{x}) = \widehat{p}(z|\mathbf{x}).\frac{d\beta(\alpha)}{d\alpha} \label{eq:correction}
\end{equation}
This means that our corrected PDF equals the initial PDF multiplied by a correction factor which is the local PIT distribution evaluated at the coverage corresponding to various redshifts. This relation is very similar to what \citet{Bordoloi2010PhotozCalibration} uses to re-calibrate photo-$z$ PDFs except now the correction factor is calculated using the local PIT distribution rather than the empirical distribution obtained from the calibration set as a whole.

The local CDF of PIT values ($r^{\widehat{p}}_{\alpha}$) is often noisy (see Fig.~\ref{fig:local}). So we implement the re-calibration method defined by Eq.~\ref{eq:correction}, with a smooth version of $\beta(\alpha)$ and its derivative. Since, $\beta(\alpha)$ is the CDF corresponding to the PIT distribution, it should be a monotonic non-decreasing function of $\alpha$. Therefore, we use a basis of $I$-spline functions to fit a smooth model using non-negative least square regression \citep{Ramsay1988Spline}. $I$-splines are monotonic functions, a linear combination of which with non-negative coefficients gives us any arbitrary monotonic non-decreasing function. We use splines of order 3 and use a basis of 5 splines. The number of basis splines to use is a  hyper-parameter and controls the level of smoothing. The derivative of $I$-splines can be obtained analytically ($M$-splines), which gives us a smooth representation of the correction factor when evaluated on a grid of $\alpha$ corresponding to the discrete grid of $z$ on which $\widehat{p}$ has been evaluated.

\section{Results and Discussion}
\begin{figure}
  \begin{subfigure}[b]{0.4\textwidth}
    \includegraphics[width=\textwidth]{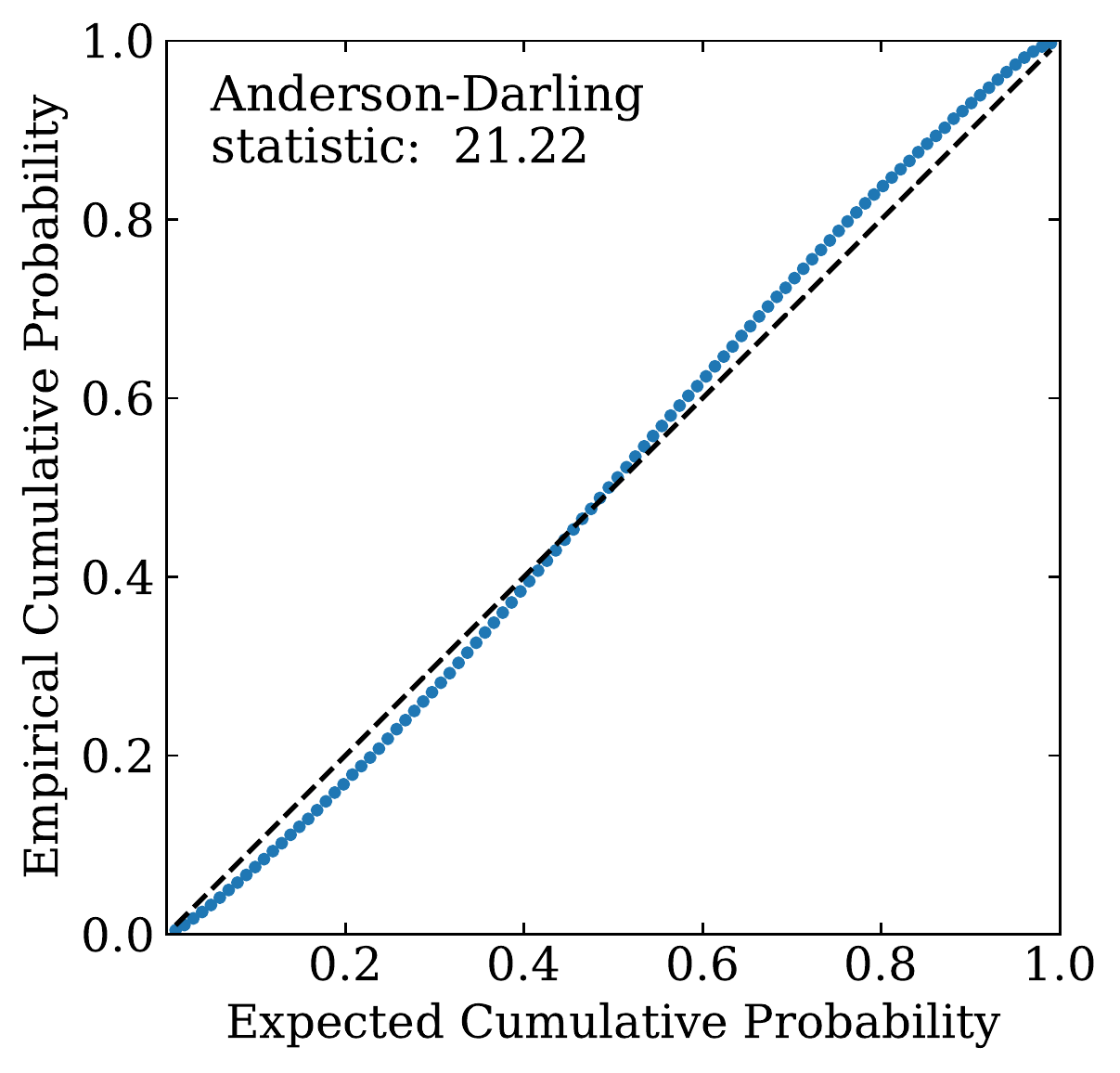}
    \caption{Original}
    \label{fig:global_uncal}
  \end{subfigure}
\hskip 3cm
  \begin{subfigure}[b]{0.4\textwidth}
    \includegraphics[width=\textwidth]{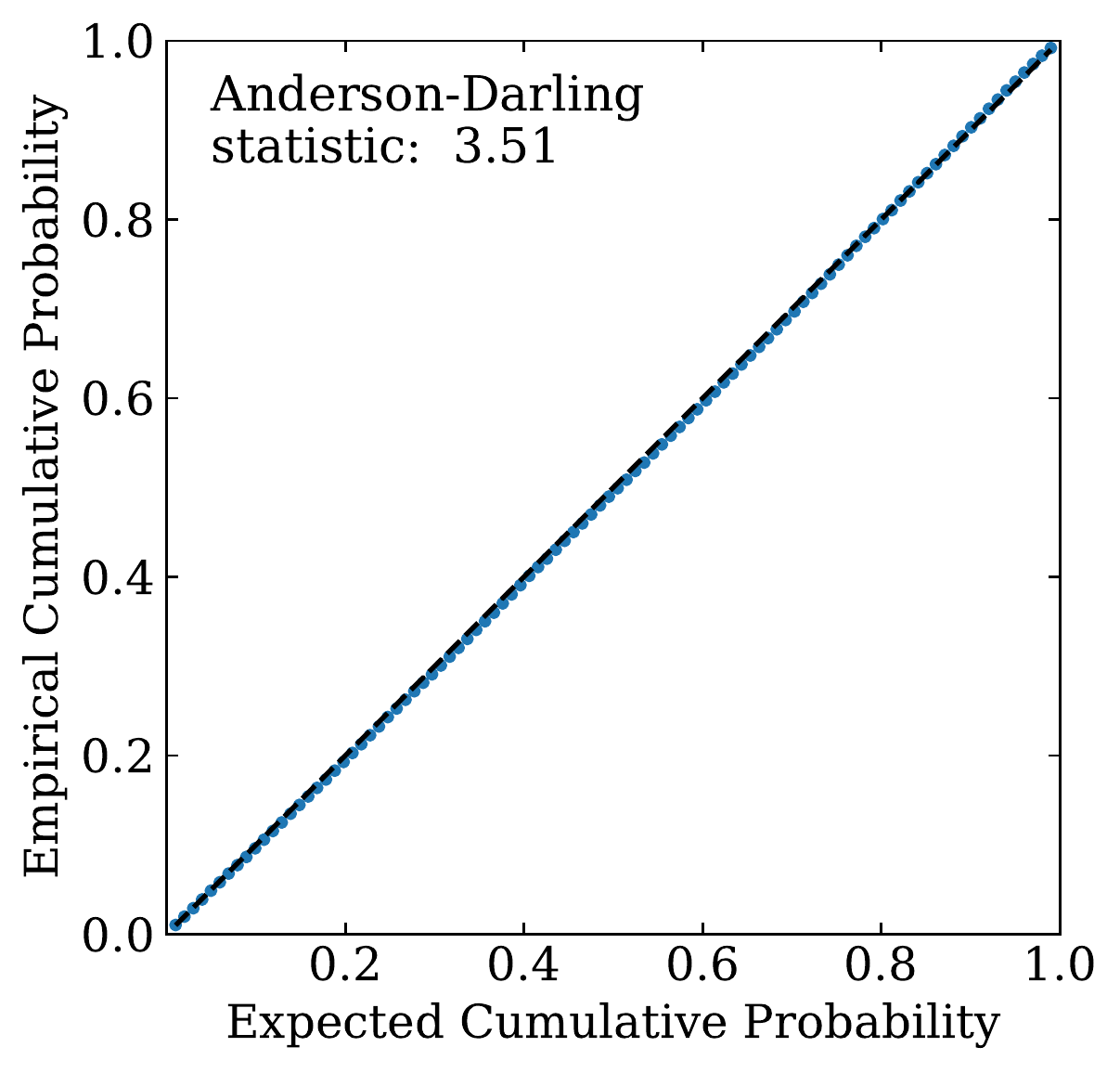}
    \caption{Re-calibrated}
    \label{fig:global_cal}
  \end{subfigure}
  \caption{P-P plot of the global distribution of PIT values for the test sample (TEDDY-C). The blue dots show the empirical CDF of the PIT values calculated as a function of their theoretical expected value. Ideally the empirical CDF and the theoretical CDF should be equal and follow the identity line (black dashed line). Fig.~\ref{fig:global_uncal} shows the P-P plot for the PDFs predicted by FlexZBoost and Fig.~\ref{fig:global_cal} shows the P-P plot for the same data set after local re-calibration. We see that after re-calibration empirical and expected CDFs match closely which is also quantified by a significant decrease in the value of the Anderson-Darling Statistic.}
\label{fig:global}
\end{figure}

\begin{figure}
 
    \includegraphics[width=\textwidth]{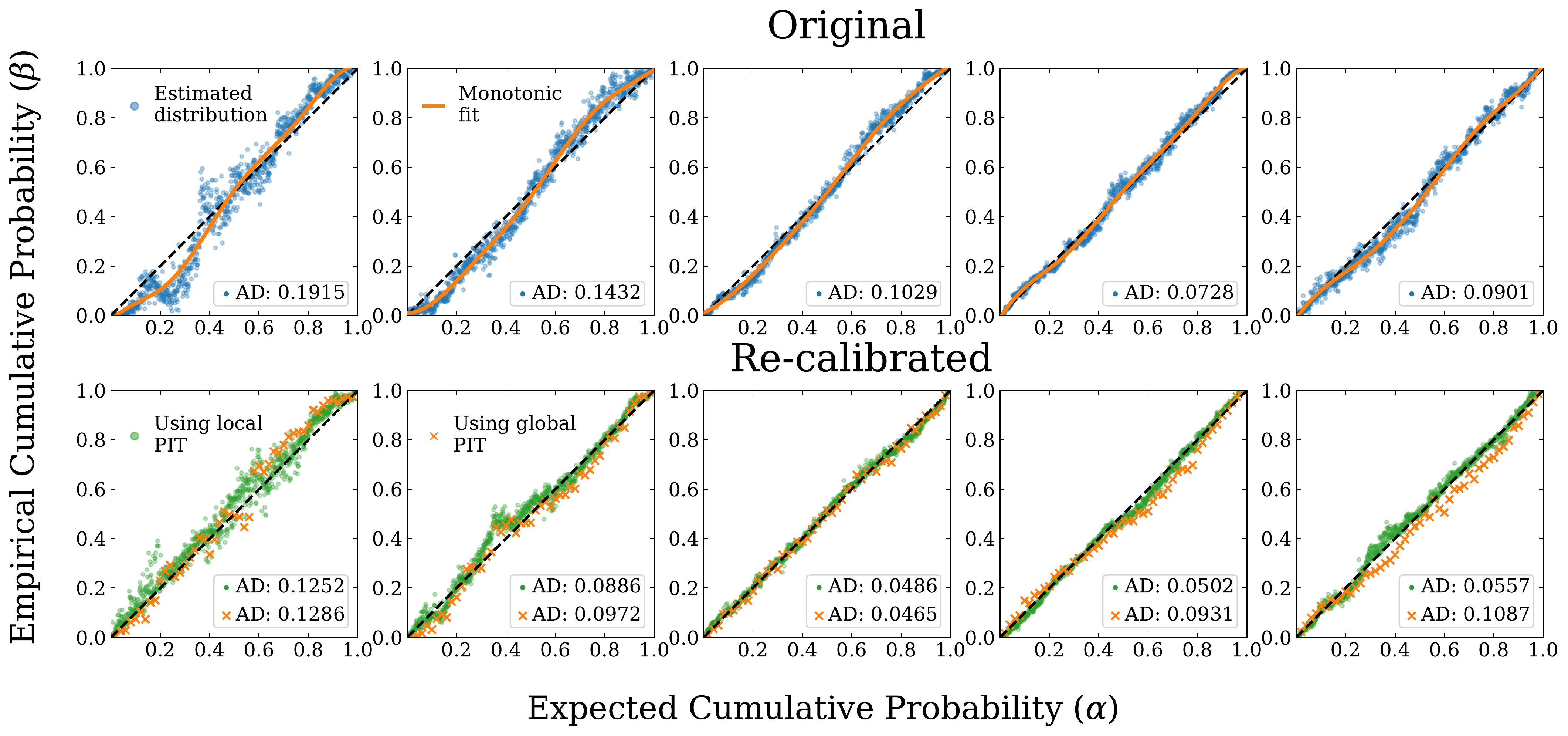}
    \caption{P-P plot of the local distribution of PIT values for 5 randomly selected objects from the test sample (TEDDY-C). The blue dots show the empirical CDF for the PIT distribution estimated as a function of their theoretical expected value using the method described in \citet{Zhao2021CDEDiagnostics}. Ideally the empirical CDF and the theoretical CDF should be equal and follow the identity line (black dashed line). The orange curve in top row shows the (least squares) fit to the blue dots using a basis of 5 $I$-splines. This fitted model is used to calculate the correction factor from eq.~\ref{eq:correction}. The top row shows the local P-P plot for the original set of PDFs, whereas the bottom row shows the local P-P plot for the globally (orange crosses) and locally (green dots) re-calibrated PDFs for the same set of objects. After re-calibration, the local P-P plots are closer to the identity line and have a lower value of the Anderson-Darling (AD) statistic. The re-calibration done using the local PIT distribution tends to perform better than the re-calibration done using the global distribution of PIT values, as seen from the lower value of the AD statistic for a majority of the cases.}
    \label{fig:local}
  
\end{figure}

We apply our method to re-calibrate PDFs obtained from FlexZBoost on both the TEDDY-B and TEDDY-C data sets and observe improvements in calibration. In addition to using P-P plots for a visual comparison, we use the Anderson-Darling (AD) statistic which is a weighted mean-squared difference between the theoretical and empirical CDFs of the PIT distributions. A lower value will indicate that the two CDFs are more similar. For the sake of brevity, we present the results from TEDDY-C as it is a more challenging scenario given the difference in the distributions of the calibration and test sets. Fig.~\ref{fig:global} shows the P-P plot for the global distribution of PIT values for the entire TEDDY-C test set before and after the re-calibration procedure. The initial P-P plot (Fig.~\ref{fig:global_uncal}) deviates from the identity line but after re-calibrating using the local distribution of PIT values follows the identity line closely (Fig.~\ref{fig:global_cal}), which is also evident from the significant decrease in the value of the AD statistic. Local calibration is a stronger requirement than global calibration so we expect that PDFs that are well calibrated locally will also be well calibrated globally.

Fig.~\ref{fig:local} shows the local P-P plots for a random subset of galaxies from TEDDY-C. The top row shows the local distribution of PIT values as inferred using a regression method trained on the calibration set and a smooth representation of the same obtained by fitting a monotonic function to the data using a basis of $I$-spline functions. After the re-calibration procedure was applied (both using global and local PIT distributions) to the entire TEDDY-C data set, we use half of TEDDY-C to train the regression model on the re-calibrated PDFs to predict the CDF of PITs for the other half of the data set. The bottom row shows a comparison of the local P-P plots re-calibrated using both local and global distribution of PIT values. We see that the P-P plots follow the identity line more closely after global re-calibration and perform even better with local re-calibration. This is again evident from the decrease in the value of the AD statistic after re-calibration. We find that the local re-calibration method tends to outperform the global re-calibration in most cases.

This work shows that PDFs can be re-calibrated using local information and produce better uncertainty estimates. Though the method works reasonably well when the distribution of features for the calibration set is slightly different from the test set, we expect performance to worsen if the distribution of features for the test set is drastically different. A systematic study to understand the performance of this method for various distributions of input features will be performed in a future work.

\section*{Broader Impact}
Machine learning algorithms are being increasingly used in decision making processes, including situations where human lives are at stake like medical diagnostics and autonomous transportation. It is therefore important that the algorithms produce accurate estimates of prediction uncertainty along with their predictions, so that we know how much confidence we can place on their predictions. This works aims to improve the quality of machine learning based predictions by proposing a general purpose method to produce well calibrated uncertainty estimates. The methods developed in this work can help us make better and hopefully unbiased decisions informed by machine learning methods.

\section*{Acknowledgments}
This material is based upon work supported by the National Science Foundation (NSF) under Grant No. AST-2009251. Any opinions, findings, and conclusions or recommendations expressed in this material are those of the author(s) and do not necessarily reflect the views of the National Science Foundation.

This work was carried out in part at ``Quarks to Cosmos with AI'', a conference supported by the NSF AI Institute: Physics of the Future, NSF PHY-2020295.  This work used the Extreme Science and Engineering Discovery Environment (XSEDE; \citealt{Towns2014XSEDE}), which is supported by National Science Foundation grant number ACI-1548562. Specifically, it used the Bridges-2 system, which is supported by NSF award number ACI-1928147, at the Pittsburgh Supercomputing Center (PSC).

 Argonne National Laboratory's work was supported by the U.S. Department of Energy, Office of Science, under contract DE-AC02-06CH11357.

We would like to thank Michael Stanley and Andresa Campos for their help and comments on this work.

\bibliographystyle{plainnat}
\bibliography{bibliography}
\end{document}